# A Lung Nodule Dataset with Histopathology-based Cancer Type Annotation

## Authors
Muwei Jian[1,2*], Hongyu Chen[2#], Zaiyong Zhang[3#], Nan Yang[2#], Haorang Zhang[1], Lifu Ma[4], Wenjing Xu[2], Huixiang Zhi[2]

**Affiliations**
1. School of Computer Science and Technology, Shandong University of Finance and Economics, Jinan, China.
2. School of Information Science and Technology, Linyi University, Linyi, China.
3. Thoracic Surgery Department of Linyi Central Hospital, Linyi, China.
4. Personnel Department of Linyi Central Hospital, Linyi, China.
[#]They contribute equally to this work and share the first authorship.

corresponding author(s): Muwei Jian (jianmuwei@ouc.edu.cn)

## Abstract

Recently, Computer-Aided Diagnosis (CAD) systems have emerged as indispensable tools in clinical diagnostic workflows, significantly alleviating the burden on radiologists. Nevertheless, despite their integration into clinical settings, CAD systems encounter limitations. Specifically, while CAD systems can achieve high performance in the detection of lung nodules, they face challenges in accurately predicting multiple cancer types. This limitation can be attributed to the scarcity of publicly available datasets annotated with expert-level cancer type information. This research aims to bridge this gap by providing publicly accessible datasets and reliable tools for medical diagnosis, facilitating a finer categorization of different types of lung diseases so as to offer precise treatment recommendations. To achieve this objective, we curated a diverse dataset of lung Computed Tomography (CT) images, comprising 330 annotated nodules (nodules are labeled as bounding boxes) from 95 distinct patients. The quality of the dataset was evaluated using a variety of classical classification and detection models, and these promising results demonstrate that the dataset has a feasible application and further facilitate intelligent auxiliary diagnosis.

## Background & Summary

At present, cancer stands as one of the most challenging diseases for medical professionals worldwide. According to the statistics of the International Agency for Research on Cancer (IARC) of the World Health Organization (WHO) in 2021[1], there were 2.21 million cases of lung cancer, accounting for 11% of all type cancers. Among them, 1.8 million people died of lung cancer bearing the proportion of 81% of the total number of lung cancer patients. It's apparent that lung cancer has become the most prevalent and dangerous disease among various cancers.

Nevertheless, ratifying to see that as technology evolves, the latest report of the American Cancer Society released in 2024, the mortality rate of lung cancer in the United States has dropped by 33%[2]. This is attributed to advancements in early lung cancer detection technology and corresponding treatment modalities. Currently, imaging scanning and pathological examination serve as two primary screening methods for lung cancer in differential diagnosis. The result of pathological examination is the gold standard for lung cancer diagnosis[3]. However, it is usually necessary to take biopsies by means of puncture or surgery, which is harmful to the human body and therefore unsuitable for routine examination. Both CT and chest X-ray photography are imaging inspection techniques based on radiology[4,5], with the difference that X-ray is performed in anteroposterior or left-right overlapping views of the body, whereas CT is performed in cross-sectional views of the body. The National Lung Screening Trial (NLST) demonstrated that annual low-dose CT screening reduces lung cancer mortality in high-risk populations compared to annual chest X-ray screening[6]. This superiority can be attributed to CT's capacity to visualize the three-dimensional structure of the lungs, thereby enabling more precise assessment of lesion location, size, and density. Consequently, CT screening significantly enhances the detection rate of lung cancer[7,8]. Nevertheless, it is still challenging to detect and identify benign and malignant pulmonary nodules based on CT. During a manual examination, an experienced physician should analyse a large number of CT images comprehensively, usually taking several minutes to thoroughly diagnose a patient, resulting in a significant workload[9]. In addition, there are diverse types of nodules that differ in size, heterogeneity and texture/appearance of the lesions in CT slices. While physicians can assess the malignancy of a nodule based on its morphology, this process heavily relies on the clinical experience and expertise of the specialist, and different doctors may give distinct diagnoses and predictions[10].

Since the beginning of the new millennium, researchers have endeavored to develop Computer Aided Detection (CAD) systems to assist doctors in diagnosing diseases efficiently[11,12]. The discerning of pulmonary nodules by traditional CAD systems mainly relies on morphological operations or low-level descriptors[13-16]. Due to the diversity of nodules in various size, arbitrary shape and varying type, the results produced by the existing traditional methods are often unsatisfactory. With the continuous development of deep learning, a large number of general-purpose models such as Region-Convolutional Neural Network (RCNN[17]) have been introduced into the field of medical image analysis. Based on these models, researchers have developed many new methods for lung nodule detection tasks[18-20]. For instance, Setio et al.[21] proposed multi-view Convolutional Neural Network (CNN) for the detection of pulmonary nodules. In this method, three types of nodules, namely solid, subsolid, and large nodules, are detected independently to reduce false positives. Ding et al.[22] applied the deconvolution structure for candidate detection on axial slices in RCNN, which improved the performance of the model during detecting nodules. Dynamic scaling cross entropy and squeeze-and-excitation block was proposed by Li et al.[23] to reduce the false detection rate of nodules, and designed a 3D CNN with a codec structure for lung nodule detection, which has achieved good outcomes. Kim et al.[24] proposed a multi-scale progressive integrated CNN based on progressive feature extraction strategy is offered to learn multi-scale input features. At the same time, significant progress has been made in the classification of benign and malignant pulmonary nodules, which is closely related to the detection task, resulting in breakthroughs in this field as well[25-28].

However, the functions of most existing CAD systems are not comprehensive, as they only involve the detection of pulmonary nodules or the benign/malignant classification of pulmonary nodules, without demonstrating more advanced capabilities. We believe that this may be directly related to the types of existing public datasets, because deep learning models are heavily dependent on training datasets, especially in the field of supervised learning. Thus, the diversity of dataset labels will directly affect the ability of model to handle different tasks. Practical applications require CAD system to have more powerful functions. Specifically, in addition to detecting areas of lung nodules and differentiating between benign and malignant lung nodules, these systems should possess the ability to predict multiple lung cancer types. This function is not redundant, as different types of lung cancer have diverse diffusion capabilities, diffusion speeds, and risks. Determining the specific category of lung cancer is closely related to the follow-up treatment methods and diagnosing approaches. Unfortunately, the aforementioned issues are not addressed in the existing openly available dataset[29-34], resulting in CAD systems in clinical applications being currently limited to the adjunctive diagnosis of lung nodules.

To address the above issues, we have curated a novel lung CT dataset, distinguished by three key characteristics: **(I) Precise cancer type labeling.** Most of the samples in dataset contain cancer type labels derived from the results of the patient's clinical diagnosis, frozen diagnosis, and pathological diagnosis, and are comprehensively considered by professional doctors, making the labels more accurate. **(II) Challenging tiny nodule detection.** The dataset comprises a substantial number of CT image samples featuring nodules of tiny and small size, which presents significant challenges for the detection of lung nodules based on CT images.

**(III) Abundant categories for cancer classification.** Given that all patients are either confirmed or suspected cases of lung cancer, it is hard to identify samples of different categories in this dataset, which poses notable challenges to cancer classification.

The contributions of this article are summarized as below:

1) We have developed a novel lung CT dataset, specifically designed for lung nodule detection and cancer classification. The dataset comprises 95 sets of CT sequences and encompasses a total of 330 nodules, among which a considerable number are tiny and prove to be challenging to recognize. Consequently, these tiny and small size nodules pose a significant challenge for both industry and academic research to the detection performance of CAD systems.

2) It is worth noting that we have integrated the results of patient clinical diagnosis (n=1), frozen diagnosis (n=1), and pathological diagnosis (n=1), supplementing this with labeled lung cancer types (lung cancer types are labeled with integer characters) on 308 samples as well as containing corresponding nodules. These include 103 benign samples, 172 Adenocarcinoma (AC) samples, and 33 Squamous Cell Carcinoma (SCC) samples with varying degrees of difficulty in classification.

3) We have conducted and evaluated extensive experiments on existing detection and classical classification models using the constructed dataset. The experimental results indicate that the existing detection models do not exhibit ideal performance in detecting tiny nodules in the dataset, and there are certain difficulties in distinguishing between different types of cancer. Therefore, this dataset is highly challenging and can promote the development and improvement of CAD systems in the future.

The dataset collected in this study focuses on improving the accuracy of lung nodule classification and detection at a single time point. The main goal of this research is to provide reliable tools for medical diagnosis, allowing for a finer categorization of different types of lung diseases so as to offer more precise treatment recommendations to healthcare professionals. In contrast, the dataset in the other work [ ] emphasizes the collection of information spanning multiple time periods for lung nodules, highlighting the dynamic changes in patients throughout the dynamic progression of the disease. The aim is to provide healthcare professionals with a comprehensive perspective, supporting individualized treatment decisions as well as furnishing spatial-temporal variation and predictions for the long-term management of patients.

The two aforementioned research projects collect data in different hospitals and aim to provide two obviously distinct but complementary perspectives. To sum up, this study emphasizes the provision of tools for the optimisation of lung disease classification models, while the other work principally concentrates on understanding the dynamic progression of the disease in patients. We believe that both distinct studies complements each other and contribute to providing promising support for early intervention and personalized treatment of lung nodules from dual perspectives.

## Methods

We collaborate with Linyi Central Hospital to collect and annotate a unique lung CT scan dataset consisting of chest CT scan images of 95 patients admitted between 2019 and 2023 (36 males and 59 females; age range: 34-78 years; mean age: 56.5 years; detailed distribution shown in Fig. 1). Different from previous studies, our dataset includes specific labels for distinct types of lung cancer derived from clinical, frozen and pathological diagnostic information corresponding to the patient's CT scan. Notably, each patient received clinical diagnostic information on the same day as the CT scan, while frozen and pathological diagnostic information were usually provided within a maximum of three days thereafter. The data samples utilized in this study underwent review by the Medical Ethics Committee of Linyi Central Hospital (Review No. LCH-LW-2022025). Given that this dataset was gathered retrospectively and all sensitive patient information was desensitized using the 3D-Slicer tool (https://www.slicer.org/), thereby ensuring that the rights and health of the subjects were not adversely affected, the ethics committee waived the necessity for researchers to obtain informed consent from patients. Additionally, the ethics committee have granted permission for the dataset to be published.

This section provides a concrete overview of the data collection process and data pre-processing methods.

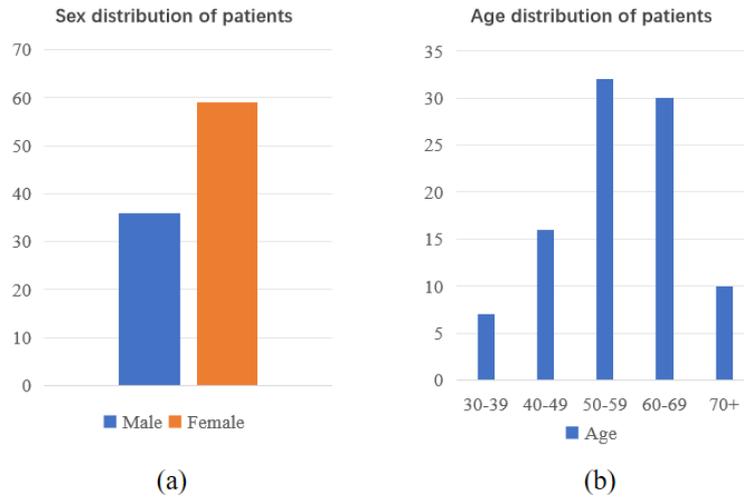

Fig. 1 Distribution of patient demographic information: (a) sex distribution; (b) age distribution.

**Data Collection and Annotation**

The CT images in the dataset were primarily sourced from Linyi Central Hospital and included preoperative CT scans, lesion locations, lesion contours and other medical information. The CT scans were obtained from a range of CT manufacturers and corresponding models (GE MEDICAL SYSTEMS Optima CT660; SIEMENS SOMATOM Definition Flash; SIEMENS Sensation 64; SIEMENS SOMATOM Force; GE MEDICAL SYSTEMS LightSpeed16; GE MEDICAL SYSTEMS BrightSpeed; UIH uCT 550) by means of different convolution kernels (B70f; B60f). Some patients underwent multiple CT scans at different time intervals. We annotate the CT scan under the guidance of professional clinicians, as shown in Fig. 2.

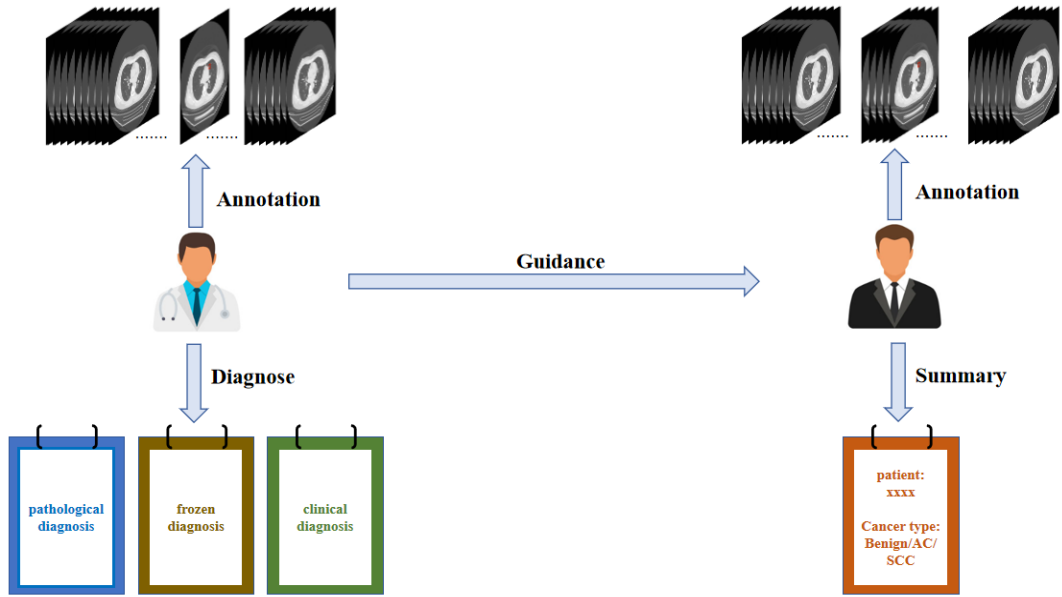

Fig. 2 The annotation process of the dataset.

In order to maintain the fidelity of annotation results, we implemented a two-stage process for annotating CT scans. In the first labelling stage, a CT scan is performed by two clinicians in the hospital who simultaneously record the image information of the lung nodule and verify the exact location of the lesion, and then generate a medical report with the corresponding clinical, frozen, and pathological diagnostic information. In case of inconsistencies between the two doctors during the first annotation phase (e.g., different lesion locations or lung cancer type diagnoses), they would discuss to determine the final outcome and form medical reports for all CT scans. In the second annotation stage, two annotators labled the position of the lung nodules on all sections of both image formats (each annotator was responsible for the annotation of one image format). They specify the maximal/minimal $x$, $y$, $z$ coordinates of the lung nodules and the diameter of the individual nodules. In the meantime, the lung cancer types from the medical reports were converted to integer character labels and assigned to the corresponding CT scans. These sections labeled with the cancer type can be further employed for prediction studies in various cancer analyses. During the second annotation stage, if the two annotators hold inconsistent annotation opinions, they will discuss and reach a final decision based on the medical reports. Throughout the annotation process, we employ anonymization technology to remove sensitive information associated with the medical images. This approach safeguards patient privacy by excluding details such as patient name, image capture time, hospital name, and attending doctor. Finally, in the CT scans of 95 patients, we annotated 330 nodules and 308 CT slices with cancer category labels. Fig. 3 provides an illustration of nodule labeling.

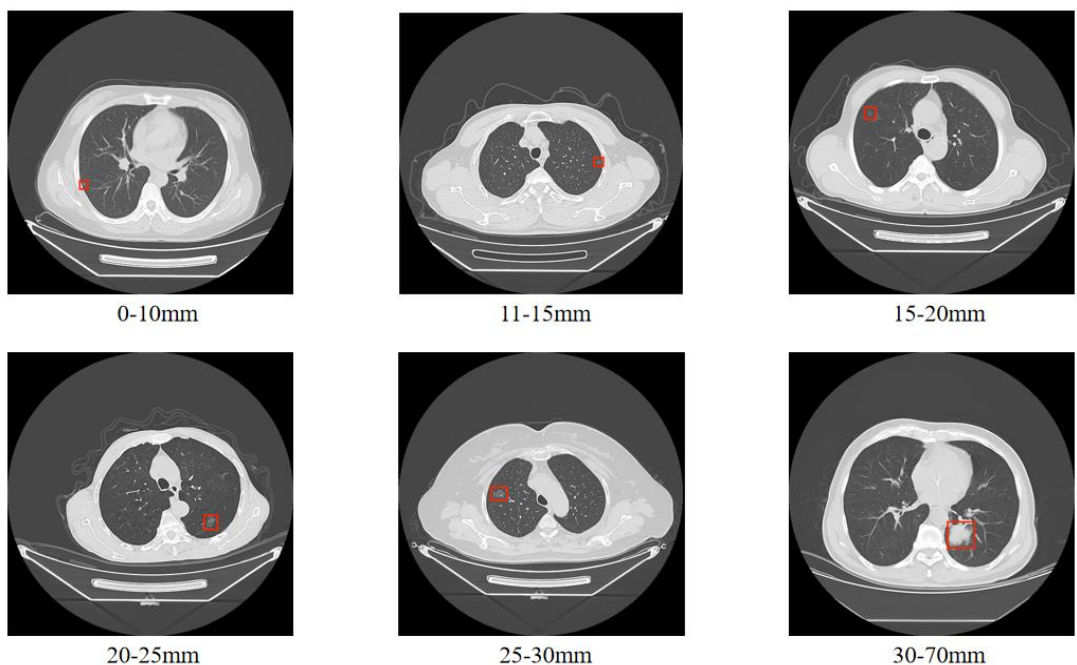

Fig. 3 Exemplars of CT slices featuring nodules with spanning various sizes.

**Data Preprocessing**

Before the comparative experiment, we carried out the following preprocessing on individual medical image:

  1) All raw data are converted to Hounsfield Units (HU).

  2) The HU value of all images is limited to the range of [-1200, 600], because the lung structure within this range is the clearest.

  3) For all samples, the image is resized into 512×512 and 36×512×512 respectively.

  4) All images are subjected to data augmentation techniques including horizontal flipping, vertical flipping, and rotation.

## Data Records

We have made the dataset publicly available in Zenodo[35,36], a public repository accessible without any password requirement. Detailed instructions for accessing the dataset are available at the following links https://doi.org/10.5281/zenodo.8422229 (version 1) and https://doi.org/10.5281/zenodo.11024613 (version 2), both are released under the Creative Commons Attribution (CC-BY) licence.

The dataset is divided into four parts. The dataset is divided into four parts, where version 1 contains bmp and mhd format data for 2D lung nodule detection, 3D lung nodule detection and cancer type prediction tasks, and version 2 includes dicom format files. Specifically, the dicom file part contains the raw data, which can be converted and used by researchers according to their needs for dicom format data. The remaining three parts are seperated based on specific task requirements (i.e., they are used for 2D lung nodule detection, 3D lung nodule detection, and cancer type prediction tasks, respectively). Fig. 4 shows the file hierarchy of the dataset with detailed descriptions as follows:

We have organized all the raw dicom data into a 'DICOM.rar' zip file. The package contains 95 subfolders named after serial numbers, each containing a CT scan.

For the 2D pulmonary nodule detection task, we have organized all samples and files within the 'BMP_2D.zip' archive. This archive comprises three subfolders: 'Image', 'Annotations', and 'ImageSets'. The 'Image' folder houses all lung CT images in BMP format; 'Annotations' contains fundamental information pertaining to 2D images, including attributes such as height, width, and depth, alongside nodule localization details. 'ImageSets' encompasses two files: 'train.txt' and 'test.txt', which serve to partition the dataset into training and testing sets adhering to a 4:1 ratio, respectively.

For the task of 3D pulmonary nodule detection, we furnish data in both BMP and MHD formats, archived respectively within 'BMP_3D.zip' and 'MHD_3D.zip'. Specifically, 'BMP_3D. zip' includes 95 subfolders named by serial numbers (such as '0001', '0002'), each containing a set of BMP format images representing CT sequence samples of a patient. In contrast, the CT sequence samples in 'MHD3D. zip' are stored in MHD and RAW formats, such as '1.mhd' and '1.raw', representing the CT sequence samples of the first patient, where MHD and RAW respectively store non-image information (image size, number of slices, etc.) and image information of the samples. Additionally, we also provide two distinct types of annotation files in CSV format. The file 'all_anno_3D.csv' encompasses nodule position data within the image coordinate system, inclusive of maximum, minimum, and central values for $x$, $y$, and $z$, as well as the diameter parameter of individual nodule size. Meanwhile, 'anno_WorldCoord.csv' incorporates nodule position information within the global coordinate system, presenting central values for $x$, $y$, and $z$, alongside nodule size details, both diameter and radius are included.

Regarding the cancer prediction investigation, BMP format images are stored in 'BMP_classification.zip'. Within this archive, the images are partitioned into training and testing sets at a ratio of 4:1 and organized into folders named 'train' and 'test', respectively. Each folder further embraces three subfolders denoted as '0', '1', and '2', wherein images belonging to distinct categories are contained. Specifically, '0', '1', and '2' correspond to 'Benign', 'ACC', and 'SCC' classifications, respectively.

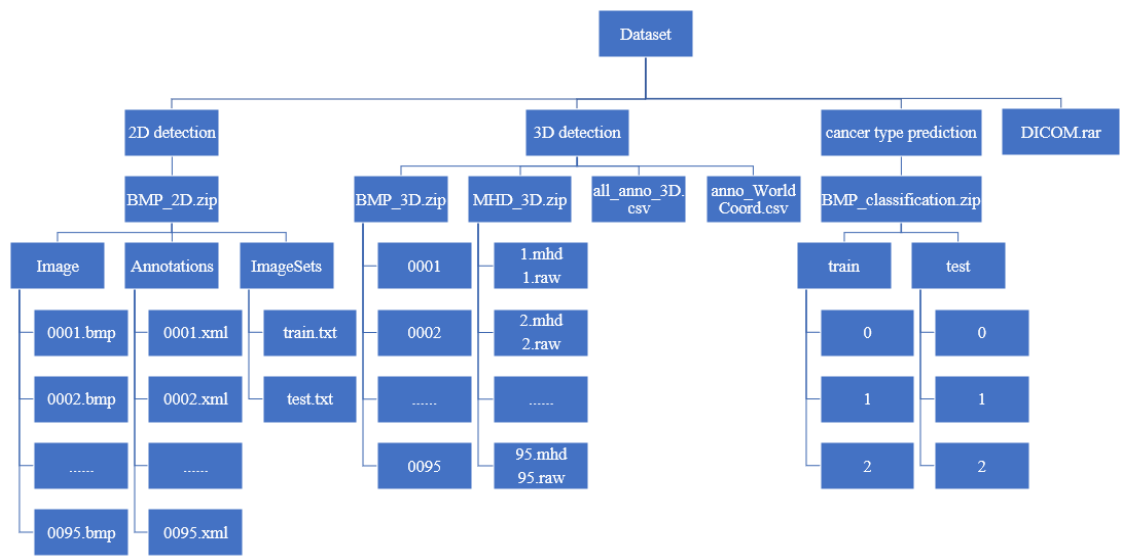

Fig. 4 Hierarchy of dataset files.

## Technical Validation

### Cancer Type Prediction

To assess the applicability and generalization of this dataset in cancer prediction and classification tasks, we conduct a comprehensive and comparative experiments employing ten representatively distinct models, including ResNet[37], EfficientNet[38], CABNet[39], ResNext[40], Res2Net[41], SE-ResNet[42], Vision-Transformer[43], InceptionV4[44], ConvNext[45], and Swin-Transformer[46]. These typical networks exhibit diverse architectural designs and characteristics, and have previously demonstrated outstanding performance in the domain of medical image classification. Concurrently, we employ two objective metrics, namely Accuracy (ACC) and Quadratic Weighted Kappa (QWK), to provide a rigorous quantitative assessment of the classification performance. Specifically, the QWK metric gauges the alignment between the outcome of individual model and the ground truth, enhancing the objectivity and impartiality of the prediction results when considered alongside the ACC indicators.

Table 1. Performance of cancer types prediction models based on the constructed dataset.

| Model | ACC | QWK |
|---|---|---|
| ResNet[37] | 0.6197 | 0.4339 |
| EfficientNet[38] | 0.6197 | 0.4583 |
| CABNet[39] | 0.5493 | 0.4861 |
| ResNext[40] | 0.5775 | 0.4548 |
| Res2Net[41] | 0.6479 | 0.3102 |
| SE-ResNet[42] | 0.6338 | 0.4750 |
| ViT-Transformer[43] | 0.6056 | 0.4468 |
| InceptionV4[44] | 0.6202 | 0.5175 |
| ConvNext[45] | 0.6056 | 0.4185 |
| Swin-Transformer[46] | 0.5352 | 0.3191 |

Table 1 presents a detailed record of the experimental results, revealing that the ACC criterion of all models is higher than 50%. This underscores the challenge of this dataset and its complicacy in practical applications, which will boost and promote further research on related medical imaging. It is also noteworthy that, in contrast to the ACC metric, the performance of the individual network in terms of the QWK evaluation indicator is also significantly lower. To delve into the underlying factors contributing to this observation, we conducted calculation and generated a confusion matrix based on the model's prediction results. The visualization results in Fig. 5, clearly depict that non-cancerous lesion and cancer samples, as well as SCC and AC samples, are extremely prone to confusion during the classification process. The reason is that samples of different categories in dataset have high similarity in image features, which leads to network confusion of indistinguishable lesion features in the course of classifying stage, thereby resulted in prediction errors. The above experimental results can prove that this dataset is a hugely challenging dataset in the field of cancer prediction. They are expected to foster future research and deeper exploration in the industry and academia, and to provide impetus for the more effective development of CAD systems.

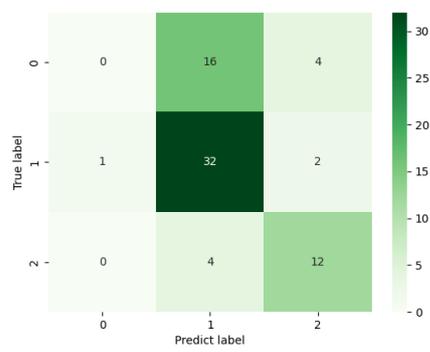

(a)

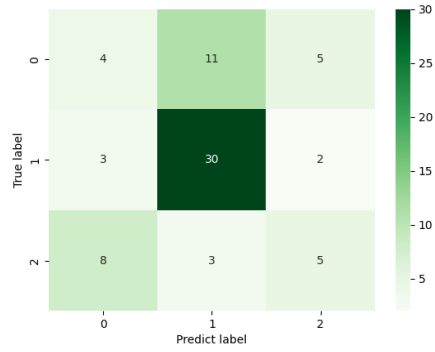

(b)

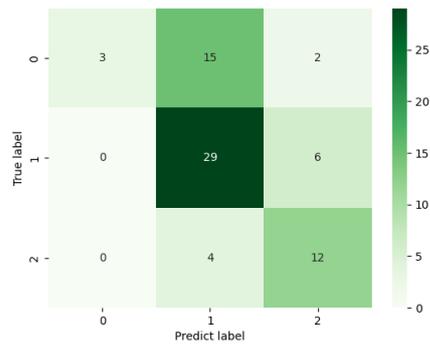

(c)

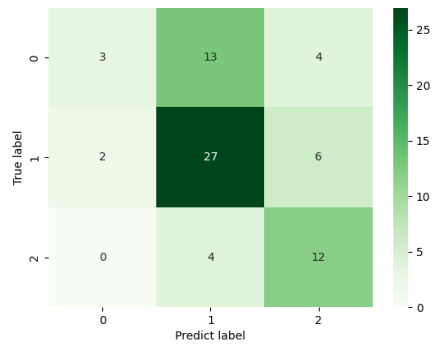

(d)

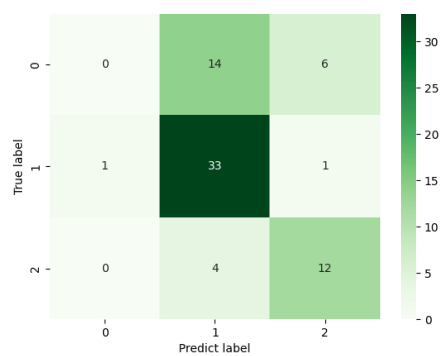

(e)

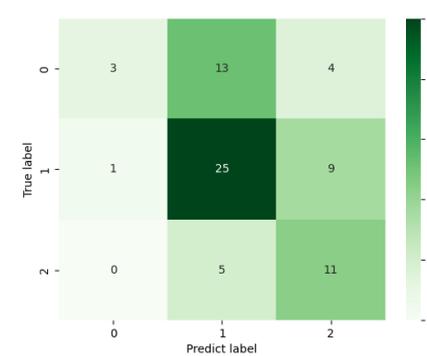

(f)

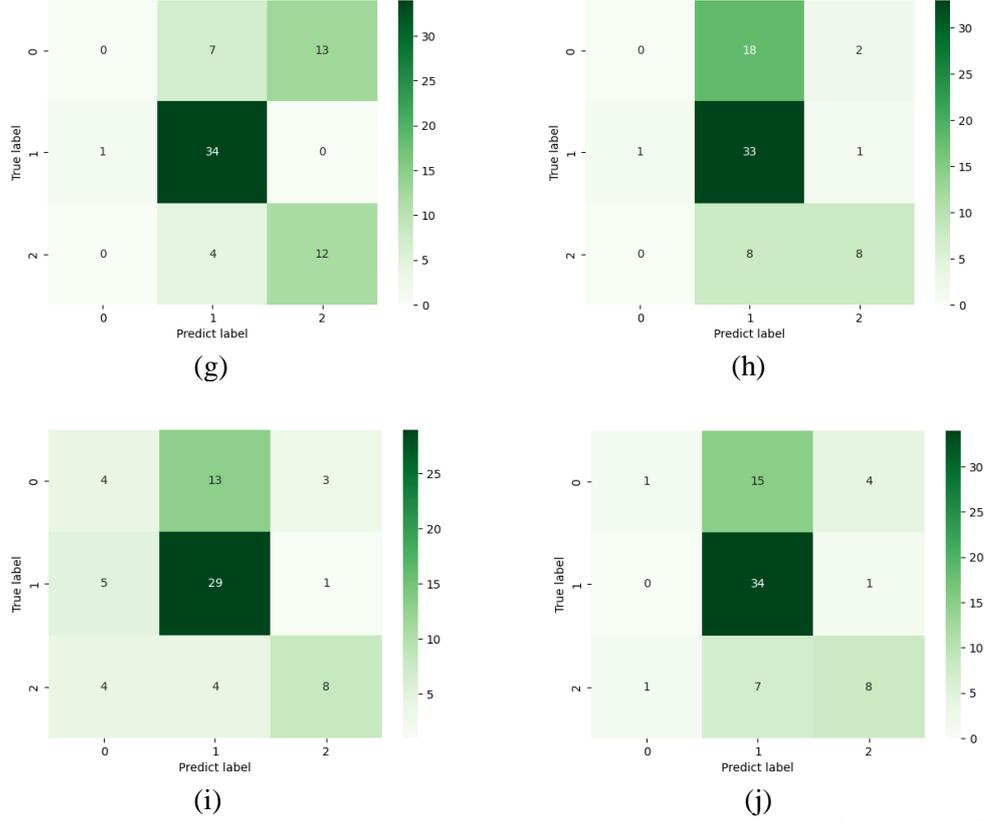

Fig. 5 The confusion matrix image of distinct classification model. (a) ResNet; (b) CABNet; (c) EfficientNet; (d) ConvNext; (e) SE-ResNet; (f) ResNext; (g) Res2Net; (h) Vision-Transformer; (i) Swin-Transformer; (j) InceptionV4.

**Pulmonary Nodule Detection**

In order to verify the effectiveness of dataset in lung nodule detection tasks, we have conducted tests on five different states of the art models, all of which are efficient approaches in the field of medical object detection, including Faster RCNN[47], Yolo[48], MobileNet[49], SSD[50], and RetinaNet[51]. In terms of evaluation indicators, we use Average Precision (AP) and Average Recall (AR) as the evaluation basis, and derived more rigorous qualitative metrics by modifying the threshold, involving $AP_{50}$, $AP_{75}$, $AP_S$, $AP_M$, $AR_1$, $AR_{10}$, $AR_S$, and $AR_L$.

Table 2. Evaluation results of different detection models based on the constructed dataset, where $AP_{50}/AP_{75}$ denotes AP at IoU (Intersection over Union) =0.5/0.75, $AP_S/AP_M$ and $AR_S/AR_L$ are AP for small/medium objects and AR for small/medium objects, respectively. $AR_1/AR_{10}$ represents AR given 1/10 detection per image.

| Model | AP | $AP_{50}$ | $AP_{75}$ | $AP_S$ | $AP_M$ | $AR_1$ | $AR_{10}$ | $AR_S$ | $AR_M$ |
|---|---|---|---|---|---|---|---|---|---|
| MobileNet[49] | 0.4136 | 0.8479 | 0.3209 | 0.4087 | 0.8515 | 0.5194 | 0.5209 | 0.5182 | 0.8500 |
| Faster R-CNN[47] | 0.5429 | 0.9272 | 0.5357 | 0.5383 | 0.9010 | 0.6463 | 0.6567 | 0.6545 | 0.9000 |
| SSD[50] | 0.5574 | 0.8894 | 0.5547 | 0.5569 | 0.8515 | 0.6179 | 0.6254 | 0.6242 | 0.8500 |
| RetinaNet[51] | 0.5807 | 0.9144 | 0.5635 | 0.5774 | 0.9010 | 0.6657 | 0.6716 | 0.6697 | 0.9500 |
| Yolo[48] | 0.6424 | 0.9388 | 0.7694 | 0.6672 | 0.9505 | 0.7284 | 0.7507 | 0.7485 | 0.9500 |

As shown in Table 2, it is evident that various models have demonstrated commendable detection performance on the dataset, reflecting the good usability of the dataset in the field of lung nodule detection. Throughout the experiment, we observed a noteworthy discrepancy between $AP_{50}$ and $AP_{75}$ metrics. Specifically, we noted that the $AP_{75}$ scores were markedly lower when compared to other evaluation indicators. This phenomenon can be attributed to the inclusion of diversely tiny nodules that are tough to observe and distinguish in the medical image, suffering that this may cause the model hard to accurately locate the nodules despite being able to detect their presence. Similarly, the results of $AP_S/AR_S$ are significantly lower than those of $AP_M/AR_M$, reinforcing the ascertainment that nodule size has a direct impact on detection results, with smaller nodules being more difficult to detect. For a more intuitive validation, we visualised the prediction results of the model with the highest average evaluation metrics (i.e. Yolo) in Table 2, as shown in Fig. 6. By observation, it can be seen that the prediction box for large nodules is almost no different from the ground-truth, but for small nodules, the model may have misdiagnosed normal lung tissue as pulmonary nodules, as well as the prediction box deviating too much from the ground-truth. Through the above analysis, we believe that this dataset can facilitate the improvement of the ability of future CAD systems to detect small and tiny pulmonary nodules, and will provide researchers with more data support for algorithm design and evaluation.

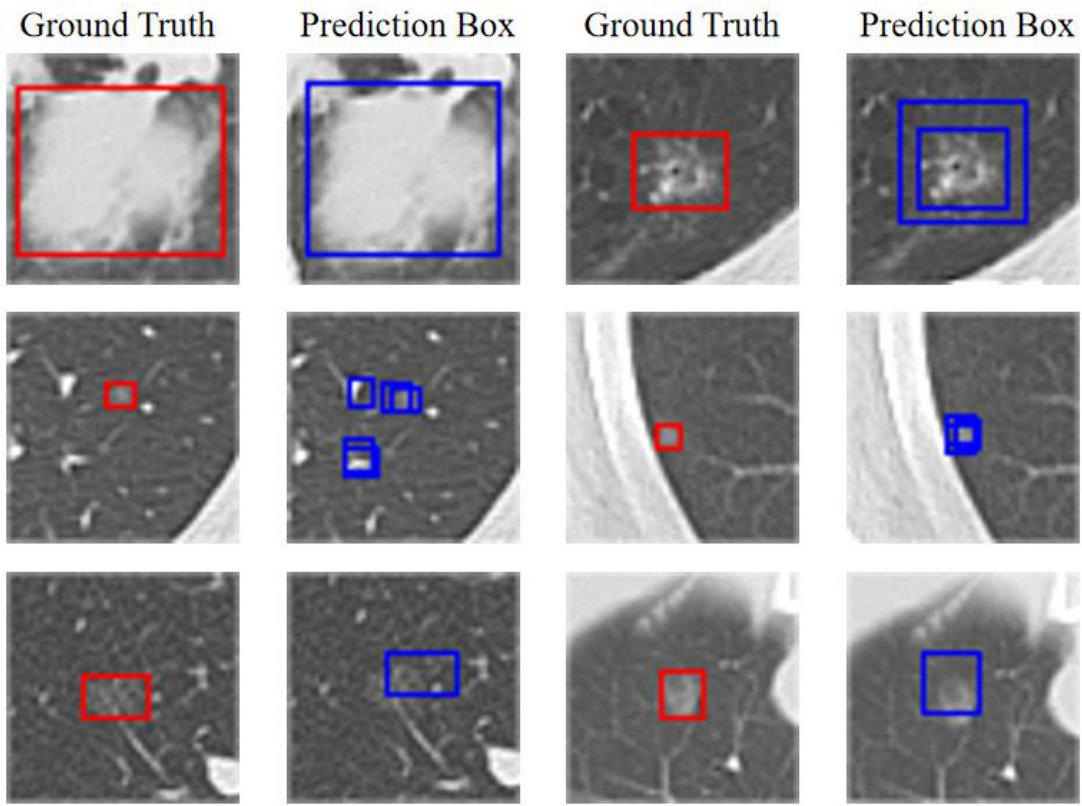

Fig. 6 Model visualisation results, where red boxes indicate ground truth and blue boxes indicate prediction boxes.

## Code Availability

In order to enhance accessibility of this dataset for users, we have made it available through a dedicated GitHub repository. This repository hosts Python sample code for data type conversion and data manipulation, aimed at facilitating researchers' comprehension and utilization of the dataset. The resource can be freely accessed at the following GitHub repository: https://github.com/chycxyzd/LDFC.

## Acknowledgements

This work was supported by National Natural Science Foundation of China (NSFC) (61976123, 61601427); Taishan Young Scholars Program of Shandong Province; and Key Development Program for Basic Research of Shandong Province (ZR2020ZD44).

## Author contributions

Muwei Jian: Writing - Conceptualization; Methodology; Writing - Review & Editing; Supervision;
Hongyu Chen: Formal analysis; Data Curation; Writing - Original Draft; Writing - Conceptualization;
Zaiyong Zhang: Methodology; Formal analysis; Resources;
Nan Yang: Validation; Conceptualization; Visualization; Data Curation;
Haoran Zhang: Validation; Data Curation;
Lifu Ma: Resources; Data Curation;
Wenjing Xu: Visualization; Data Curation;
Huixiang Zhi: Visualization; Data Curation.

## Competing interests

The authors declare that they have no competing interests.